\documentclass[aps,graphicx,amsmath,twocolumn]{revtex4}

\usepackage{graphicx}

\begin{document}


\title{Self-error-rejecting quantum state transmission of entangled photons for faithful quantum communication without calibrate reference frames}

\author{Peng-Liang Guo,$^{1,*}$  Tao Li,$^{2,1,\footnote{The first two authors contributed
equally to this work.}}$   Qing Ai,$^{1}$
 and  Fu-Guo
Deng$^{1,
}$\footnote{Corresponding author: fgdeng@bnu.edu.cn} }

\address{$^{1}$Department of Physics, Applied Optics Beijing Area Major Laboratory,
Beijing Normal University, Beijing 100875, China\\
$^{2}$ School of Science, Nanjing University of Science and Technology,
Nanjing 210094, China
}

\date{\today }

\begin{abstract}
We propose an alignment-free two-party polarization-entanglement transmission scheme for entangled photons by using only linear-optical elements, requiring neither ancillary photons nor calibrated reference frames. The scheme is robust against both the random channel noise and the instability of reference frames, and it is subsequently extended to multi-party Greenberger-Horne-Zeilinger state transmission. Furthermore, the success probabilities for two- and multi-party entanglement transmission are, in principle, improved to unity when active polarization controllers are used. The distinct characters of a simple structure, easy to be implemented, and a high fidelity and efficiency  make our protocol very useful for long-distance quantum communications and distributed quantum networks in practical applications.
\end{abstract}

\maketitle

\section{Introduction}

Quantum entanglement is an important resource for quantum information science~\cite{QC,qc2}. The correlations between entangled systems constitute an essential building block that enables various quantum networking and quantum computing, such as quantum key distribution (QKD)~\cite{BB84,B92,EK91,QKD1}, quantum
teleportation~\cite{TELE1,TELE2,TELE3}, quantum dense coding~\cite{dense1}, quantum secret sharing~\cite{QSS1,QSS2,QSS3}, quantum secure direct communication~\cite{QSDC1,twostep,twostepexp,twostepexp2}, and distributed quantum computation~\cite{qc1,BorregaardPRL2015,Qindistributed1,Sheng2017,Qindistributed2}. A prerequisite to all these quantum tasks is to establish a faithful entanglement channel between authorized parties.   It is   known that the quantum entanglement can only be created locally and be distributed to distant communicating parties through practical quantum channels (an optical-fibre channel or a free-space one)~\cite{QC,qc2}.
Photons are natural flying qubits for carrying quantum information as they are fast and weak to interact to their environment. They can be transmitted from one location to another through fibre or free-space channels, which enables the shared entanglement between distant parities. Especially, the polarization degree of freedom (DOF) of single photons, with clear practical advantages, is extensively exploited to complete various kinds of quantum information tasks~\cite{KokRMP,Dengphoton}.

When photons pass through fibre or free-space channels in practice, they will inevitably suffer from channel noise, such as the fibre birefringence or the
atmospheric turbulence, leading to the decoherence of the photons~\cite{Kimble2008,Dur1999,Sheng2013}. The decoherence will degrade the fidelity of entanglement shared between the parties. Moreover,  even for a noiseless channel with ideal photon transmission, the misalignment of reference frames between the communicating parties will directly affect their measurements relative to each other~\cite{RFmis0,RFmis1,Andreoli2017}, thus it decreases the  entanglement available for post-processing quantum information tasks~\cite{QC,qc2,BB84,B92,EK91,QKD1,TELE1,TELE2,TELE3,dense1,QSS1,QSS2,QSS3,QSDC1,twostep,twostepexp,twostepexp2}.
For instance, in the polarization encoding, the horizontal and vertical polarization states constitute a basis for encoding a qubit~\cite{KokRMP,Dengphoton}, which can be denoted as
$\left|0\right\rangle\rightarrow\left|H\right\rangle$ and
$\left|1\right\rangle\rightarrow\left|V\right\rangle$. It is well
known that the horizontal (H) and vertical (V) polarization states are
defined with respect to a certain reference frame. Unfortunately, the reference
frames of the distant parties usually are not stable and possibly change
over time, due to the random influence introduced by environments. It is therefore hard for the communicating parties to align their reference frames perfectly~\cite{RFsource2001,RFsource2004}. The photon, prepared in a pure state $|\Psi\rangle$ relative to the reference frame of the sender, will be described by $\hat{U}(\theta)|\Psi\rangle$ with respect to the reference frame of the receiver, where $\hat{U}(\theta)$ represents a random rotation of $\theta$ between two reference frames of the parties~\cite{RFmis0}. This partly distorts the original photonic state, and reduces the security and efficiency of practical quantum communicating networks~\cite{QC,qc2,BB84,B92,EK91,QKD1,TELE1,TELE2,TELE3,dense1,QSDC1,QSS1,QSS2,QSS3,twostep,twostepexp,twostepexp2}.

So far, there are some interesting protocols for quantum state transmission and entanglement distribution. The earliest protocols exploit quantum-error-correcting codes to rectify either bit- or phase-flip errors~\cite{qecc,BFER}. The redundant encoding logical qubits, after passing through noisy channels, can be retrieved by post-selecting measurement and quantum feedback. Then, a much more efficient method is developed by resorting to decoherence-free subspaces (DFSs)~\cite{DFS}, and encodes logical qubits in a DFS that protects the qubits against channel noise~\cite{DFS1,DFS2,DFS3,DFS4,Yamamoto,DFS6,Li2008}. Usually, it requires the use of two or more photons to encode one logical qubit in a DFS. In 2003, Walton \emph{et al}.~\cite{DFS2} proposed a QKD protocol by encoding  each decoherence-free qubit in the time and phase of a pair of photons. Subsequently, Boileau \emph{et al}.~\cite{DFS3} proposed a robust quantum cryptography protocol with a DFS constituted by polarization-entangled photon pairs. In 2007,  Yamamoto \emph{et al}.~\cite{Yamamoto} presented a scheme for single-photon state transmission by assisting an additional photon, and then performed  parity-check measurements~\cite{parity}
and posted-selection to overcome collective-noise effects. The scheme is subsequently extended to polarization entanglement distribution,
by assisting more single photons~\cite{DFS6}.
These schemes usually need two or more photons to encode a logical qubit. Furthermore, multipartite encoding protocols make the quantum state transmission more susceptible to photon losses, and require that photons from each logical qubit suffer from exactly the same noise for an ideal transmission, which is hard to achieve, especially for long-distance quantum communications, using the existing technology~\cite{RFmis0}.

An alternative DFS-based quantum state transmission, without a shared reference frame, could be constituted by encoding a logical qubit with multiple DOFs of a single photon~\cite{ET, mdf1,mdf2,mdf3,Vallone2014}. The net effect of the channel noise, applied on multiple DOFs encoded photons, could be sufficiently suppressed when its effect on each DOF cancels each other completely. For example, a logical qubit, defined by the polarization along with transverse-spatial-mode~(TSM)~\cite{mdf2} or orbital-angular-momentum~(OAM)~\cite{mdf3} states, can be decoherence free in the paraxial approximation. The quality of TSM
and the stability of OAM play a key role in these schemes, while it is hard to transmit quantum states encoded in TSM or OAM over hundreds meters currently~\cite{mdf2,Vallone2014}. Furthermore, a liquid crystal device \emph{q}-plate~\cite{qp1,qp2} is needed to map a polarization-encoded qubit into a  hybrid polarization-OAM-encoded logical qubit, and vice versa. This significantly limits the efficiency of these schemes, due to the low performance of current \emph{q}-plates.

In this article, we propose a high-fidelity
polarization-entanglement transmission scheme with a new encoding-decoding process, and it requires neither ancillary photons nor calibrated reference frames. The decoherence effect introduced by the misalignment of reference frames can be completely suppressed by
linear optical elements and post selections.
The parties in quantum communication can share the desired polarization entanglement with a success efficiency that depends on random rotation angles between their reference frames. Furthermore, the success efficiency is, in principle, improved to near-unity by assisting active polarization modulators or Pockels cells~(PCs). Moreover, our protocols for two-party quantum communication are general and they can be directly extended to transmit Greenberger-Horne-Zeilinger~(GHZ) polarization entanglement for multi-party quantum communicating networks.   These distinct features of a simple structure, easy to be implemented, and a high fidelity and efficiency  make this protocol very useful for long-distance quantum communication, quantum networks, and distributed quantum computation in practical applications in the future.

\begin{figure}[!tpb]
\centering
\includegraphics[width=6.6cm,angle=0]{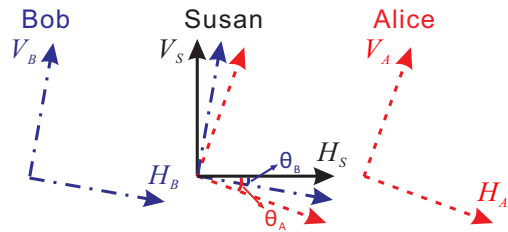}
\caption{The misalignment of reference frames of receivers
Alice and Bob versus the reference frame of the sender Susan.
$\theta_{A}$ and $\theta_{B}$ are the angles between the reference frames of Alice and Bob relative to the reference frame of Susan.} \label{fig1}
\end{figure}

\begin{figure}[!tpb] 
\begin{center}
\includegraphics[width=8.0 cm,angle=0]{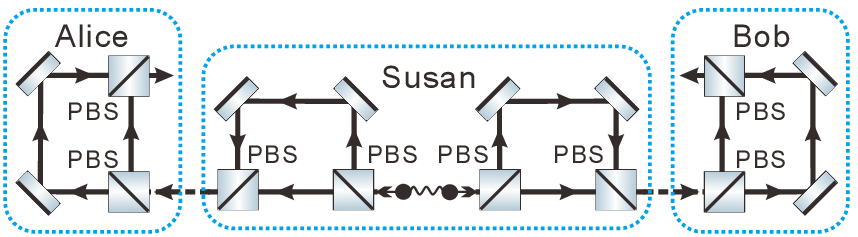}
\caption{The schematic setup for the two-party polarization-entanglement transmission
without the experimental reference-frame  alignment.  PBSs represent polarization beam splitters which transmit horizonal polarized photons $|H\rangle$ and reflect vertical polarized photons $|V\rangle$. The slim rectangles represent mirrors with a unity reflection coefficient. \label{fig2}}
\end{center}
\end{figure}

\section{Entangled photon-pair transmission without an alignment} \label{Sec-2}

Suppose the sender Susan, located in the middle of two communicating parties Alice and Bob, produces photon pairs in a polarization-entangled state in her reference frame, and this state can be written as
\begin{eqnarray}
\left|\psi\right\rangle_{S}=\alpha\left|H\right\rangle_{S}\left|H\right\rangle_{S}
+\beta\left|V\right\rangle_{S}\left|V\right\rangle_{S}.\label{entangle1}
\end{eqnarray}
Here $\alpha$ and $\beta$ are two arbitrary
parameters with $\left|\alpha\right|^2+\left|\beta\right|^2=1$. $\left|H\right\rangle$ and $\left|V\right\rangle$ represent the
horizontal and vertical polarization states of single
photons in the corresponding reference frame, respectively. The subscript $S$
denotes that the state here is relative to the reference frame of Susan.
When the two parties Alice and Bob do not share the same reference frame to that of Susan,  there are two unknown misalignment angles $\theta_{A}$ and
$\theta_{B}$ between the two reference frames of Alice and Bob and that of Susan, shown
in Fig.~\ref{fig1}. The blue dotted lines, the black solid lines, and the red dashed lines represent rectangular axes in reference frames of Bob, Susan, and Alice, respectively. Moreover, these angles are random and  vary in time, and it is an experimental challenge to eliminate the effect of the misalignments by active feedback rotations~\cite{RFsource2001,RFsource2004}.

The polarization states relative to the reference frame of Susan will be redescribed relative to the reference frame of Alice (Bob) as follows:
\begin{equation}
\begin{split}
\left|H\right\rangle_{S}&=\cos\theta_{i}\left|H\right\rangle_{i}
+\sin\theta_{i}\left|V\right\rangle_{i}, \\
\left|V\right\rangle_{S}&=-\sin\theta_{i}\left|H\right\rangle_{i}+\cos\theta_{i}\left|V
\right\rangle_{i}.
\end{split}
\end{equation}
Here the subscript $i=A$ or $B$, and it denotes that the state is described relative to the reference frame of Alice or Bob, respectively. Therefore, the misalignment between reference frames of communicating parties can be regarded as an unknown collective-rotation noise.

The entangled photon pair, shared by
Alice and Bob, deviate significantly to their original state, shown in Eq.(\ref{entangle1}), due to the misalignment of the reference frames. The photon-pair state can be redescribed relative to reference frames of Alice and Bob as:
\begin{equation}
\begin{split}
   \left| \psi\right\rangle ^{d}_{AB}\!=& \cos\theta _{A}' \cos\theta _{B}' \left ( \alpha \left | H \right\rangle_{A}\left | H \right\rangle_{B}+\beta  \left | V\right\rangle_{A}\left | V\right\rangle_{B} \right) \\
   &+\cos\theta _{A}'\sin \theta _{B}' \left ( \alpha \left | H \right\rangle_{A}\left | V \right\rangle_{B}-\beta  \left | V\right\rangle_{A}\left | H\right\rangle_{B} \right ) \\
   &+\sin\theta _{A}' \cos \theta _{B}' \left ( \alpha \left | V \right\rangle_{A}\left | H \right\rangle_{B}-\beta  \left | H\right\rangle_{A}\left | V\right\rangle_{B} \right ) \\
   &+\sin\theta _{A}' \sin\theta _{B}' \left ( \alpha \left | V \right\rangle_{A}\left | V \right\rangle_{B}+\beta  \left | H\right\rangle_{A}\left | H\right\rangle_{B} \right).\;\;\;\;
\end{split}
\end{equation}
Here $\theta_{A}'$ and $\theta_{B}'$ represent effective rotation angles, which  is a combined effect originating from the channel rotation and the reference frame misalignment noises.
The superscript $d$ denotes that the state is obtained by direct transmission without using our scheme. The subscripts $A$ and $B$ denote that the polarization of the photon, owned by Alice and Bob, are described in Alice's and Bob's experimental reference frames, respectively.
The fidelity of the state $\left| \psi\right\rangle ^{d}_{AB}$, with respect to the original entangled state $\left|\psi\right\rangle_{S}$, is $F_1=|_{AB}^{d}\langle\psi|\psi\rangle_{S}|^2$ and it can be detailed as \begin{eqnarray}
F_{1}=|\cos\theta _{A}' \cos\theta _{B}'+\sin\theta _{A}' \sin\theta _{B}'(\beta^{*}\alpha+\alpha^{*}\beta)|^{2},
\end{eqnarray}
which depends on the random rotation angles $\theta _{A}'$ and $\theta _{B}'$.
Therefore, the direct transmission, without an alignment, is unfaithful, and it will reduce the security and efficiency of practical quantum communicating networks.

To overcome this problem and provide Alice and Bob with the
desired entanglement, a new time-tag encoding-decoding scheme is used here.
First, Susan passes each photon of an entangled photon pair through an unbalanced polarization interferometer, shown in Fig.~\ref{fig2}, which is composed of two polarization beam splitters (PBSs) and two high-reflective mirrors. The vertical polarization component $\left|V\right\rangle$ of each photon is thus delayed by a time scale of $T$. With this encoding process, the original entangled state  $\left|\psi\right\rangle_{S}$, sending to Alice and Bob, becomes
\begin{eqnarray}
{{\left| \psi  \right\rangle }_{S}}
\to\alpha {{\left| H \right\rangle }_{S}}{{\left| H \right\rangle  }_{S}}
+\beta {{\left| {{V}_{T}} \right\rangle }_{S}}{{\left| {{V}_{T}} \right\rangle }_{S}},
\end{eqnarray}
where the subscript $T$ denotes the delay, and the state can be viewed as a two-photon four-qubit GHZ state that is encoded in both the polarization and time-bin DOFs. Taking the channel noise and the misalignment into account, the state of the photon pair shared by Alice and Bob, after introducing an additional time-delay $T$ on the horizonal component of each photon, evolves into
\begin{equation}
\begin{split}
{{\left| \psi  \right\rangle }_{AB}}\!=&
\cos{{\theta}_{A}'}\cos{{\theta}_{B}'}\left(\alpha{{\left| {H_{T}}\right\rangle}_{A}}{{\left| {H_{T}} \right\rangle }_{B}}\!+\!\beta {{\left| {V_{T}} \right\rangle }_{A}}{{\left| {V_{T}} \right\rangle}_{B}}\right) \\
&\!+\!\cos{{\theta}_{A}'}\sin{{\theta}_{B}'}\left(\alpha{{\left|{H_{T}}\right\rangle}_{A}}{{\left|V\right\rangle}_{B}}\!-\!\beta{{\left| {V_{T}}\right\rangle}_{A}}{{\left|{H_{TT}}\right\rangle}_{B}}\right)\\
&\!+\!\sin {{\theta}_{A}'}\cos{{\theta}_{B}'}\left(\alpha{{\left|V\right\rangle}_{A}}{{\left|{{H}_{T}}\right\rangle}_{B}}\!-\!\beta {{\left|{{H}_{TT}}\right\rangle }_{A}}{{\left|{{V}_{T}}\right\rangle}_{B}} \right)\\
&\!+\!\sin{{\theta}_{A}'}\sin{{\theta}_{B}'}\left(\alpha {{\left|V\right\rangle}_{A}}{{\left| V\right\rangle}_{B}}\!+\!\beta {{\left| {{H}_{TT}} \right\rangle }_{A}}{{\left| {{H}_{TT}} \right\rangle}_{B}}\right).\;\;
\end{split}
\end{equation}
Here the subscript $T$ and $TT$ denote a single and double delays,
respectively. The original entangled state $\left|\psi\right\rangle_{S}$ can be
picked out with unity fidelity by selecting photon pairs with a delay $T$ on each photon.
The success probability (or the efficiency) for this transmission is
\begin{eqnarray}
 P_1=\cos^2\theta_{A}'\cos^2\theta_{B}',
\end{eqnarray}
and it is dependent on the value of $\theta_{A}'$ and $\theta_{B}'$, which changes slightly during the transmission process.

\begin{figure}[!tpb] 
\begin{center}
\includegraphics[width=7.2cm,angle=0]{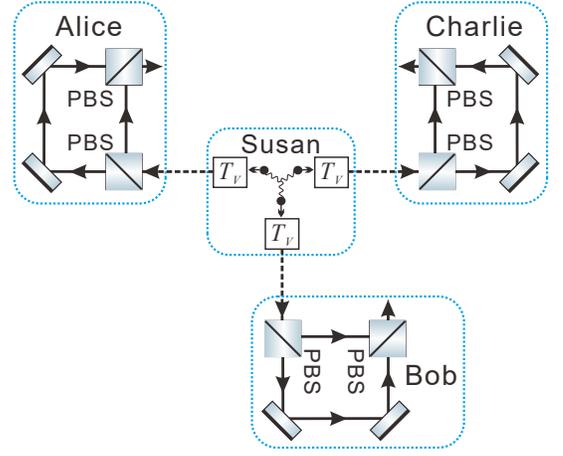}
\caption{The schematic setup for the alignment-free transmission of three-photon GHZ polarization entanglement with linear optical elements. $T_V$ stands for a time-tag operation and completes a time-delay of $T$ on the vertical polarization component of photons passing through it.}\label{fig3}
\end{center}
\end{figure}

\section{ GHZ-entanglement triplet transmission without an alignment} \label{Sec-3}

The principle of our alignment-free entanglement transmission scheme for
two legitimate parties can be extended to the multi-participant
case directly. For example, in a three-party case, a three-photon polarization-entanglement source locates at the node of Susan among three distant parties, Alice, Bob, and Charlie. It produces  polarization-entangled photon triplets in a GHZ state~\cite{ghz1,ghz2},
which can be written as
\begin{eqnarray}
{{\left| \phi  \right\rangle }_{S}}
=\alpha {{\left| H \right\rangle }_{S}}{{\left| H \right\rangle }_{S}}
{{\left| H \right\rangle }_{S}}+\beta {{\left| V \right\rangle }_{S}}
{{\left| V \right\rangle }_{S}}{{\left| V \right\rangle }_{S}}.\label{entangle2}
\end{eqnarray}
Here the subscript $\emph{S}$ denote these states are described relative to the reference frame
of Susan. She sends each photon of an entangled triplet, respectively,  to
three distant parties without aligned reference frames.
Similarly, there exist unknown angles ${{\theta }_{A}}$, ${{\theta }_{B}}$, and ${{\theta
}_{C}}$ varying in time between the reference frames of Alice, Bob,
and Charlie relative to the reference frame of Susan, respectively.

When the entangled photons are received, respectively, by
Alice, Bob, and Charlie, the state of the photon triplet is redescribed relative to their experimental reference frames, and it is
\begin{equation}
\begin{split}
{{\left| \phi  \right\rangle }\!_{ABC}^{d}}\!=&
\lambda_{1}\left( \alpha {{\left|{{H}} \right\rangle }\!_{A}} {{\left| {{H}}  \right\rangle}_{B}}{{\left| {{H}} \right\rangle }_{C}} \!+\! \beta {{\left|{{V}} \right\rangle}\!_{A}} {{\left| {{V}} \right\rangle
}_{B}}{{\left| {{V}} \right\rangle }_{C}} \right)  \\
 \!&+\!\lambda_{2}\left(\alpha{{\left|{{H}}\right\rangle}\!_{A}}{{\left|{{H}}\right\rangle}_{B}}{{\left|V\right\rangle}_{C}}\!-\!\beta {{\left|{{V}} \right\rangle }\!_{A}}{{\left| {{V}}\right\rangle }_{B}}{{\left| {{H}} \right\rangle }_{C}} \right)   \\
\!&+\!\lambda_{3}\left(\alpha{{\left|{{H}}\right\rangle}\!_{A}}{{\left|V\right\rangle}_{B}}{{\left|{{H}}\right\rangle }_{C}}\!-\!\beta {{\left|{{V}}\right\rangle}\!_{A}}{{\left|{{H}}\right\rangle}_{B}}{{\left|{{V}}\right\rangle}_{C}}\right) \\
 \!&+\!\lambda_{4}\left(\alpha {{\left|{{H}}\right\rangle}\!_{A}}{{\left|V\right\rangle}_{B}}{{\left|V\right\rangle }_{C}} \!+\!\beta {{\left|{{V}}\right\rangle}\!_{A}}{{\left|{{H}}\right\rangle}_{B}}{{\left|{{H}}\right\rangle}_{C}}\right) \\
 \!&+\!\lambda_{5}\left(\alpha{{\left|V\right\rangle}\!_{A}}{{\left|{{H}}\right\rangle}_{B}}{{\left|{{H}}\right\rangle}_{C}}\!-\!\beta
 {{\left| {{H}} \right\rangle }\!_{A}}{{\left| {{V}}\right\rangle }_{B}}{{\left|{{V}}\right\rangle }_{C}}\right) \\
 \!&+\!\lambda_{6}\left(\alpha {{\left| V \right\rangle}\!_{A}}{{\left|{{H}}\right\rangle}_{B}}{{\left| V\right\rangle}_{C}}\!+\!\beta {{\left| {{H}}  \right\rangle }\!_{A}}{{\left| {{V}} \right\rangle }_{B}}{{\left| {{H}} \right\rangle}_{C}}\right)  \\
 \!&+\!\lambda_{7}\left(\alpha{{\left|V\right\rangle}\!_{A}}{{\left|V\right\rangle}_{B}}{{\left|{{H}}\right\rangle}_{C}}\!+\!\beta
 {{\left|{{H}}\right\rangle }\!_{A}}{{\left| {{H}}\right\rangle }_{B}}{{\left| {{V}} \right\rangle }_{C}}  \right) \\
 \!&+\!\lambda_{8}\left( \alpha {{\left| V \right\rangle    }\!_{A}}{{\left| V \right\rangle }_{B}}{{\left| V \right\rangle }_{C}}\!-\!\beta {{\left| {{H}}\right\rangle }\!_{A}}{{\left| {{H}} \right\rangle}_{B}}{{\left| {{H}} \right\rangle }_{C}} \right),\;\;\;\;
\end{split}
\end{equation}
The coefficients are detailed as follows:
\begin{equation}
\begin{split}
\lambda_{1}=\;&\cos{{\theta'}\!_{A}} \cos{{\theta'}\!_{B}} \cos{{\theta'}\!_{C}},\\ \lambda_{2}=\;&\cos{{\theta'}\!_{A}} \cos{{\theta'}\!_{B}} \sin {{\theta '}\!_{C}},\\
\lambda_{3}=\;&\cos{{\theta '}\!_{A}} \sin{{\theta' }\!_{B}} \cos {{\theta'}\!_{C}},\\
\lambda_{4}=\;&\cos{{\theta '}\!_{A}} \sin{{\theta' }\!_{B}} \sin {{\theta'}\!_{C}},\\
\lambda_{5}=\;&\sin{{\theta }'\!_{A}} \cos {{\theta'}\!_{B}} \cos {{\theta' }\!_{C}},\\
\lambda_{6}=\;&\sin{{\theta }'\!_{A}} \cos {{\theta'}\!_{B}} \sin {{\theta' }\!_{C}},\\
\lambda_{7}=\;&\sin{{\theta }'\!_{A}} \sin {{\theta'}\!_{B}} \cos {{\theta' }\!_{C}},\\  \lambda_{8}=\;&\sin{{\theta }'\!_{A}} \sin {{\theta'}\!_{B}} \sin {{\theta' }\!_{C}}.
\end{split}
\end{equation}
Here ${\theta '}\!_{i}$ ($i=A$, $B$, or $C$) represents the net rotation angle suffered from the channel rotation noise and the reference frame misalignment, when the photons are sent to Alice, Bob, and Charlie, respectively. Therefore, the parties cannot share the desired three-photon GHZ state due to lacking of a
shared Cartesian reference frame.

The fidelity of the state ${\left|\phi\right\rangle}\!_{ABC}^{d}$, with respect to the original state ${\left| \phi  \right\rangle }_{S}$, shown in Eq.(\ref{entangle2}), is decreased to
\begin{eqnarray}
F_2=|\lambda_{1}+\lambda_{8}(\beta^{*}\alpha-\alpha^{*}\beta)|^{2}.
\end{eqnarray}
This low fidelity, in principle, can be increased to unity with a similar encoding-decoding scheme discussed above. First, Susan sends each of the three entangled photons
into an unbalanced polarization interferometer, as shown in Fig. \ref{fig3}, to complete a tag-operation $T_{V}$ and
introduce a delay of $T$ on the vertical-polarization component of each photon. Then the quantum state of the photon triplet sent to Alice, Bob and Charlie evolves into
\begin{eqnarray}
{{\left| \phi  \right\rangle }_{S}}\to\alpha {{\left| H \right\rangle }_{S}}
{{\left| H \right\rangle }_{S}}{{\left| H \right\rangle }_{S}}
+\beta {{\left| {{V}_{T}} \right\rangle }_{S}}{{\left| {{V}_{T}}
  \right\rangle }_{S}}{{\left| {{V}_{T}} \right\rangle }_{S}}.
\end{eqnarray}
When the entangled photon triplet are received by Alice, Bob and
Charlie, they evolve into a new state due to the channel noise and the misalignment of the experimental reference frames of the distant parties.

To obtain the desired entangled state, the parties perform a decoding process by applying a tag-operation $T_{H}$ and introducing a delay $T$ on the horizontal-polarization component of each photon. Now, the
three-photon state relative to the experimental reference frames of receivers becomes
\begin{equation}
\begin{split}
{{\left| \phi  \right\rangle }\!_{ABC}}\!=&
\lambda_{1}\!\left( \alpha {{\left|{{H}_{T}} \right\rangle }\!_{A}} {{\left| {{H}_{T}}  \right\rangle}\!_{B}}{{\left| {{H}_{T}} \right\rangle }\!_{C}} \!+\! \beta {{\left|{{V}_{T}} \right\rangle}\!_{A}} {{\left| {{V}_{T}} \right\rangle}\!_{B}}{{\left| {{V}_{T}} \right\rangle }\!_{C}} \right) \\
\!&+\!\lambda_{2}\!\left( \alpha {{\left| {{H}_{T}} \right\rangle }\!_{A}}{{\left| {{H}_{T}} \right\rangle}\!_{B}}{{\left| V \right\rangle }\!_{C}}\!-\!\beta {{\left|{{V}_{T}} \right\rangle }\!_{A}}{{\left| {{V}_{T}}\right\rangle }\!_{B}}{{\left| {{H}_{TT}} \right\rangle }\!_{C}} \right) \\
\!&+\!\lambda_{3}\!\left( \alpha {{\left| {{H}_{T}}\right\rangle }\!_{A}}{{\left| V \right\rangle }\!_{B}}{{\left| {{H}_{T}} \right\rangle }\!_{C}}\!-\!\beta {{\left|{{V}_{T}}\right\rangle }\!_{A}}{{\left| {{H}_{TT}} \right\rangle }\!_{B}} {{\left| {{V}_{T}} \right\rangle }\!_{C}} \right)  \\
\!&+\!\lambda_{4}\!\left( \alpha {{\left| {{H}_{T}} \right\rangle }\!_{A}}{{\left| V \right\rangle }\!_{B}}{{\left| V \right\rangle }\!_{C}}\!+\!\beta {{\left| {{V}_{T}} \right\rangle }\!_{A}}{{\left| {{H}_{TT}}\right\rangle }\!_{B}}{{\left| {{H}_{TT}} \right\rangle }\!_{C}} \right)  \\
\!&+\! \lambda_{5} \!\left( \alpha {{\left| V \right\rangle }\!_{A}}{{\left| {{H}_{T}}\right\rangle }\!_{B}}{{\left| {{H}_{T}} \right\rangle }\!_{C}}\!-\!\beta {{\left| {{H}_{TT}} \right\rangle }\!_{A}}{{\left| {{V}_{T}}\right\rangle }\!_{B}}{{\left| {{V}_{T}} \right\rangle }\!_{C}} \right)    \\
\!&+\!\lambda_{6}\!\left( \alpha {{\left| V \right\rangle   }\!_{A}}{{\left| {{H}_{T}}\right\rangle }\!_{B}}{{\left| V \right\rangle }\!_{C}}\!+\!\beta {{\left| {{H}_{TT}}    \right\rangle }\!_{A}}{{\left| {{V}_{T}} \right\rangle }\!_{B}}{{\left| {{H}_{TT}} \right\rangle }\!_{C}}  \right)   \\
\!&+\!\lambda_{7}\!\left( \alpha {{\left| V \right\rangle    }\!_{A}}{{\left| V \right\rangle }\!_{B}}{{\left| {{H}_{T}} \right\rangle }\!_{C}}\!+\!\beta {{\left| {{H}_{TT}}\right\rangle }\!_{A}}{{\left| {{H}_{TT}}
  \right\rangle }\!_{B}}{{\left| {{V}_{T}} \right\rangle }\!_{C}}  \right)\\
 \!&+\!\lambda_{8}\!\left( \alpha {{\left| V \right\rangle    }\!_{A}}{{\left| V \right\rangle }\!_{B}} {{\left| V \right\rangle }\!_{C}}\!-\!\beta {{\left| {{H}_{TT}}\right\rangle }\!_{A}}{{\left| {{H}_{TT}} \right\rangle }\!_{B}}
   {{\left| {{H}_{TT}} \right\rangle }\!_{C}} \right).\;\;\;\;\;\;\;
\end{split}
\end{equation}
Obviously, the parties can share the initial GHZ state by
picking out the component with the same delay of $T$ on each photon.
The success probability (the efficiency) of this transmission is
\begin{eqnarray}
 P_2=\cos^2\theta'_{A}\cos^2\theta'_{B}\cos^2\theta'_{C},
\end{eqnarray}
which depends on the rotation angles $\theta'_{A}$, $\theta'_{B}$, and $\theta'_{C}$.
However, once the transmission succeeds, its fidelity, in principle, approaches unity.

For a general $N$-party quantum communication, the
entangled $N$ photons are sent to $N$ receivers, say Alice, Bob, ..., and Zach, respectively. The
$N$-photon entangled state, relative to Susan's reference frame, is a GHZ-entanglement one, and it
can be written as
\begin{eqnarray}
{{\left| \varphi  \right\rangle }_{S}}=\alpha {{\left| H \right\rangle
}_{S}} {{\left| H \right\rangle }_{S}}\cdot \cdot \cdot
{{\left| H \right\rangle }_{S}}\!+\!\beta {{\left| V \right\rangle
}_{S}}{{\left| V \right\rangle }_{S}}\!\cdot \cdot \cdot\! {{\left| V
\right\rangle }_{S}}.\;\;\;\;\label{eq-n}
\end{eqnarray}
Similarly, Susan applies a tag-operation $T_{V}$ on the
vertical polarization state of each photon, and then the N-photon state becomes
\begin{eqnarray}
{{\left| \varphi \right\rangle }_{S}}\!\to\!\alpha {{\left| H
\right\rangle }\!_{S}} {{\left| H \right\rangle }\!_{S}}\cdot \cdot
\cdot {{\left| H \right\rangle }\!_{S}} \!+\!\beta {{\left| {{V}_{T}}
\right\rangle }\!_{S}}{{\left| {{V}_{T}} \right\rangle }\!_{S}}\cdot
\cdot \cdot {{\left| {{V}_{T}} \right\rangle }\!_{S}}.\;\;\;\;
\end{eqnarray}
After passing through noisy channels, the quantum state of the $N$ photons changes significantly. To recover the original GHZ entanglement, the parities apply a tag-operation $T_{H}$ and introduce a time delay of $T$ on the horizontal
polarization component of each photon. The quantum state of the $N$-photon system now evolves into
\begin{equation}
\begin{split}
{{\left| \varphi  \right\rangle
}_{AB...Z}}\!=\!&\sum_{j,k,...l}{{{p}^A_{j}}{{p}^B_{k}}...{{p}^Z_{l}}}\Big[
\alpha {{\left| {{m}_{j}}\right\rangle }_{A}}{{\left| {{m}_{k}}
\right\rangle }_{B}}\cdot \cdot \cdot {{\left| {{m}_{l}}
\right\rangle }_{Z}}  \\ 
&+ {{\left( -1 \right)}^{j+k...+l}}\beta {{\left|
{{n}_{j}} \right\rangle }_{A}}{{\left| {{n}_{k}} \right\rangle
}_{B}}\cdot \cdot \cdot {{\left| {{n}_{l}} \right\rangle }_{Z}}\Big],\;\;
\end{split}
\end{equation}
where the subscripts $i= A$, $B$, ..., and $Z$ represent photons that are described relative to the reference frames of Alice, Bob, ..., Zach, respectively;
$j,k,$ ..., $l~\in~\{0,1\}$;
${p^i_{0}}=cos{\theta' _{i}}$; ${{p}^i_{1}} =sin{{\theta' }_{i}}$;
$m_{0}=H_{T}$;
$\emph{m}_{1} =\emph{V}$;
$\emph{n}_{0}=\emph{V}_{T}$;
$\emph{n}_{1}=\emph{H}_{\emph{TT}}$.
When $j=0,k=0,$ ..., and $l=0$, there is a delay $T$ on both components of each photon. By picking out this case, the $N$ parties share the original GHZ state
$\left| \varphi  \right\rangle_{S}$ in Eq.(\ref{eq-n}).
The corresponding  success probability equals
\begin{eqnarray}
 P_3=\cos^2\theta'_{A}\cos^2\theta'_{B}...\cos^2\theta'_{Z}.
\end{eqnarray}
In addition, the fidelity of this transmission of the $N$-photon GHZ states, in principle, approaches unity. This can significantly reduce quantum sources consumed for quantum communication networks, compared to the quantum communication with a low-fidelity transmission.

\section{Alignment-free entanglement transmission with active polarization modulators} 

So far, the alignment-free entanglement transmission is achieved faithfully with only passive linear optical elements. Its fidelity approaches near-unity, at the expense of a decrease of the success probability (the efficiency). In this section, we propose a modified entanglement transmission scheme, without aligned reference frames between distant parties. The success probability is largely increased, and it can in principle approach unity, by using active polarization modulators, such as Pockels cells (PCs). The PCs rotate the polarization of photons, when a voltage is
applied to the PCs, using a fast electro-optic effect.

\begin{figure}[!tpb] 
\begin{center}
\includegraphics[width=8.5cm,angle=0]{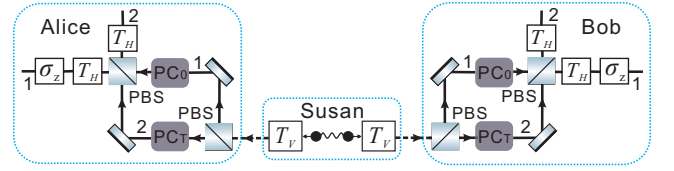}
\caption{The alignment-free setup for the two-party entanglement transmission with active
Pockels cells (PCs). PC$_0$ and PC$_T$ are activated during time slots with \emph{zero} and \emph{T} delay, respectively. $\sigma_{z}$ completes a phase-flip operation on photons passing through it.}\label{fig4}
\end{center}
\end{figure}

The schematic setup is shown in Fig.~\ref{fig4}.  Photon pairs generated by Susan are entangled in the state $\left|\psi\right\rangle_{S}$, shown in Eq.(\ref{entangle1}). Before sending a photon pair into noisy channels, Susan performs an encoding  by applying a tag-operation
$T_{V}$ and introducing a time delay $T$ on the vertical polarization component of each photon. The encoded photon pair received by Alice and Bob, after passing though noisy channels, can be redescribed relative to the experimental reference frames of Alice and Bob as
\begin{eqnarray}
{{\left|\psi'\right\rangle }_{AB}}
=&\!\!\!\!\!\!\cos{{\theta' }\!_{A}}\cos{{\theta'}\!_{B}}\left(\alpha{{\left|H\right\rangle}\!_{A}}{{\left|H \right\rangle}\!_{B}}+\beta {{\left| {{V}_{T}} \right\rangle }\!_{A}}{{\left|{{V}_{T}} \right\rangle }\!_{B}} \right)\nonumber \;\;\;\;\;\\
&\!\!\!\!\!\!+ \cos{{\theta' }\!_{A}}\sin {{\theta' }\!_{B}}\left( \alpha {{\left| H\right\rangle }\!_{A}}{{\left| V\right\rangle }\!_{B}}-\beta{{\left| {{V}_{T}}\right\rangle }\!_{A}}{{\left| {{H}_{T}} \right\rangle }\!_{B}} \right)\nonumber \;\;\;\;\;\\
 &\!\!\!\!\!\!+  \sin {{\theta' }\!_{A}}\cos {{\theta' }\!_{B}}\left( \alpha {{\left| V \right\rangle }\!_{A}}{{\left| H\right\rangle }\!_{B}}-\beta{{\left| {{H}_{T}}\right\rangle }\!_{A}}{{\left| {{V}_{T}} \right\rangle }\!_{B}}\right)\nonumber \;\;\;\;\;\\
&\!\!\!\!\!\!+\sin {{\theta' }\!_{A}}\sin {{\theta' }\!_{B}}\left( \alpha {{\left| V\right\rangle }\!_{A}}{{\left| V\right\rangle }\!_{B}}+\beta{{\left| {{H}_{T}}\right\rangle }\!_{A}}{{\left| {{H}_{T}} \right\rangle }\!_{B}}\right).\;\;\;\;\;
\end{eqnarray}
Here ${\theta' }\!_{A}$ and ${\theta' }\!_{B}$ represent effective rotation angles introduced by a combined effect originating from channel noise and  reference-frame misalignment.

To share the desired polarization entanglement, Alice and Bob perform a decoding process, respectively. They first pass their photons through a balanced interferometer that consists of two PBSs and two PCs. The PC completes a
$\sigma_{\emph{x}}=|H\rangle\langle{}V|+|V\rangle\langle{}H|$ operation on photons passing through it only when the PC is activated. Here the
$PC_{0}$ is activated only when photons without any delay passing through it, while the $PC_{T}$ is activated only when photons
with a delayed $T$ passing through it. Then each photon will be output into two possible paths. In both paths, a tag-operation $T_{H}$ is performed, and it introduces  a delay $T$ on
the horizontal polarization component of photons. Just after the tag-operation $T_{H}$, the state of the photon pair evolves into
\begin{equation}
\begin{split}
{{\left| \psi'\right\rangle }\!_{AB}}\!\to&\cos{{\theta' }\!_{A}}\cos{{\theta' }\!_{B}}\!\left( \alpha \!\left|{{H}_{T}} \right\rangle \!_{A}^{2}\left| {{H}_{T}}\right\rangle\!_{B}^{2}\!+\!\beta\! \left| {{V}_{T}}\right\rangle \!_{A}^{2}\left| {{V}_{T}}\right\rangle \!_{B}^{2} \right)\\
&+  \cos{{\theta'}\!_{A}}\sin {{\theta' }\!_{B}}\!\left( \alpha\! \left| {{H}_{T}} \right\rangle \!_{A}^{2}\left|{{H}_{T}} \right\rangle \!_{B}^{1}\!-\!\beta\! \left| {{V}_{T}} \right\rangle \!_{A}^{2}\left| {{V}_{T}}\right\rangle \!_{B}^{1} \right)\\
&+  \sin {{\theta' }\!_{A}}\cos {{\theta' }\!_{B}}\!\left( \alpha\! \left| {{H}_{T}} \right\rangle \!_{A}^{1}\left|{{H}_{T}} \right\rangle\! _{B}^{2}\!-\!\beta\! \left| {{V}_{T}}\right\rangle \!_{A}^{1}\left| {{V}_{T}}   \right\rangle \!_{B}^{2} \right)\\
&+  \sin {{\theta' }\!_{A}}\sin {{\theta' }\!_{B}}\!\left( \alpha \!\left|{{H}_{T}} \right\rangle \!_{A}^{1}\left|{{H}_{T}} \right\rangle\! _{B}^{1}\!+\!\beta\! \left| {{V}_{T}}\right\rangle \!_{A}^{1}\left| {{V}_{T}}   \right\rangle \!_{B}^{1}\right).
\;\;\;\;\;\;\;\;\;\;\;\;
\end{split}
\end{equation}
Here the superscripts 1 and 2 represent two different output paths. The two parties can share the desired entanglement sent by Susan, when the two photons are output in both path 1 or path 2. For other cases, these is a phase-flip error on the entangled photons pair. To eliminate the path-dependent phase-flip error, a phase-flip operation, ${{\sigma}_{z}}=\left| H \right\rangle \left\langle  H \right|-\left| V\right\rangle \left\langle  V \right|$, is introduced in  path 1 in each node. Now, the state of the photon pair evolves into
\begin{equation}
\begin{split}
{{\left| \psi '' \right\rangle }\!_{AB}}\!=&\cos{{\theta' }\!_{A}}\cos{{\theta' }\!_{B}}\!\left( \alpha \!\left|
  {{H}_{T}} \right\rangle \!_{A}^{2}\left| {{H}_{T}}
  \right\rangle\!_{B}^{2}\!+\!\beta\! \left| {{V}_{T}}
  \right\rangle \!_{A}^{2}\left| {{V}_{T}}
  \right\rangle \!_{B}^{2} \right)\\
  &+  \cos{{\theta'}\!_{A}}\sin {{\theta' }\!_{B}}
  \!\left( \alpha\! \left| {{H}_{T}} \right\rangle \!_{A}^{2}\left|
 {{H}_{T}} \right\rangle \!_{B}^{1}\!+\!\beta\! \left| {{V}_{T}}
 \right\rangle \!_{A}^{2}\left| {{V}_{T}}\right\rangle \!_{B}^{1} \right)\\
 &+  \sin {{\theta' }\!_{A}}\cos {{\theta' }\!_{B}}
 \!\left( \alpha\! \left| {{H}_{T}} \right\rangle \!_{A}^{1}\left|
 {{H}_{T}} \right\rangle\! _{B}^{2}\!+\!\beta\! \left| {{V}_{T}}
 \right\rangle \!_{A}^{1}\left| {{V}_{T}}   \right\rangle \!_{B}^{2} \right)\\
&+  \sin {{\theta' }\!_{A}}\sin {{\theta' }\!_{B}}\!\left( \alpha \!\left|
{{H}_{T}} \right\rangle \!_{A}^{1}\left|
 {{H}_{T}} \right\rangle\! _{B}^{1}\!+\!\beta\! \left| {{V}_{T}}
\right\rangle \!_{A}^{1}\left| {{V}_{T}}   \right\rangle \!_{B}^{1}
\right).\;\;\;\;\;
\end{split}
\end{equation}
No matter in which output paths the photon pair are received, the communicating parties, Alice and Bob, can share the initial entanglement $|\psi\rangle_S$, detailed in Eq.(\ref{entangle1}), without the alignment of reference frames.
The success probability of the transmission is improved to unity, when the PCs are ideal with unity efficiency. Moreover, the corresponding fidelity also approaches unity.

\begin{figure}[!tpb]
\begin{center}
\includegraphics[width=8.5 cm,angle=0]{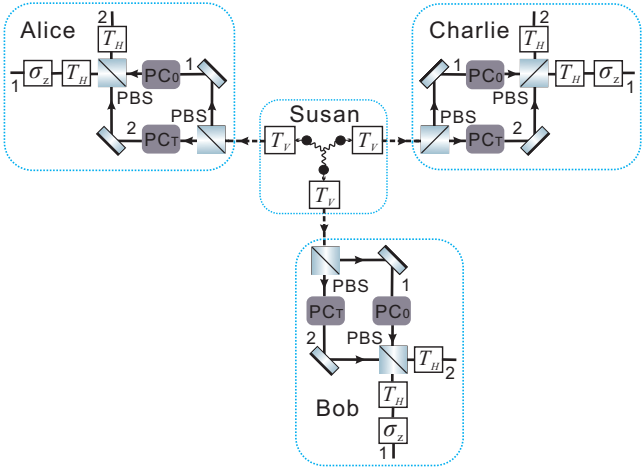}
\caption{ The alignment-free  setup for the three-photon GHZ polarization-entanglement
transmission with Pockels cells (PCs).
}\label{fig5}
\end{center}
\end{figure}

The present high-efficiency protocol can be extended to transmit a three-photon GHZ state, such as ${\left| \phi  \right\rangle }_{S}$ in Eq.(\ref{entangle2}).  A GHZ entangled  triplet first pass though an encoding circuit, which completes a tag-operation T$_V$  on the vertical polarization component of each photon. Upon arriving at the distant parties, each photon  passes though a decoding circuit, shown in Fig. \ref{fig5}. The state of the three-photon system evolves into a path-dependent one as follow:
\begin{small}
\begin{eqnarray}
{{\left| \phi'  \right\rangle }\!_{ABC}}
\!\!\!\!&=&\!\!\!\!\lambda_{1}\!\left( \alpha \left| {{H}_{T}} \right\rangle \!_{A}^{2}\left| {{H}_{T}}\right\rangle \!_{B}^{2}\left| {{H}_{T}}\right\rangle \!_{C}^{2}\!+\!\beta \left| {{V}_{T}}\right\rangle \!_{A}^{2}\left| {{V}_{T}} \right\rangle\!_{B}^{2}\left| {{V}_{T}} \right\rangle \!_{C}^{2} \right)\nonumber\\
\!\!\!\!\!&&\!\!\!\!+\!\lambda_{2}\!\left(\alpha \left| {{H}_{T}} \right\rangle\!_{A}^{2}\left| {{H}_{T}} \right\rangle \!_{B}^{2}\left| {{H}_{T}}\right\rangle \!_{C}^{1}\!+\!\beta \left|{{V}_{T}} \right\rangle \!_{A}^{2}\left| {{V}_{T}} \right\rangle\!_{B}^{2}\left| {{V}_{T}} \right\rangle\!_{C}^{1} \right)\nonumber\\
\!&&\!\!\!\!+\! \lambda_{3}\!\left(\alpha \left| {{H}_{T}} \right\rangle\!_{A}^{2}\left| {{H}_{T}} \right\rangle \!_{B}^{1}\left| {{H}_{T}}
\right\rangle\! _{C}^{2}\!+\!\beta \left|{{V}_{T}} \right\rangle \!_{A}^{2}\left| {{V}_{T}} \right\rangle\!_{B}^{1}\left| {{V}_{T}} \right\rangle\!_{C}^{2} \right)\nonumber\\
\!&&\!\!\!\!+\! \lambda_{4}\!\left(\alpha \left| {{H}_{T}} \right\rangle\!_{A}^{2}\left| {{H}_{T}} \right\rangle \!_{B}^{1}\left| {{H}_{T}}
\right\rangle \!_{C}^{1}\!+\!\beta \left|{{V}_{T}} \right\rangle \!_{A}^{2}\left| {{V}_{T}} \right\rangle\!_{B}^{1}\left| {{V}_{T}} \right\rangle\!_{C}^{1} \right)\nonumber\\
\!&&\!\!\!\!+\!\lambda_{5}\!\left(\alpha \left| {{H}_{T}} \right\rangle\!_{A}^{1}\left| {{H}_{T}} \right\rangle \!_{B}^{2}\left| {{H}_{T}}
\right\rangle \!_{C}^{2}\!+\!\beta \left|{{V}_{T}} \right\rangle \!_{A}^{1}\left| {{V}_{T}} \right\rangle\!_{B}^{2}\left| {{V}_{T}} \right\rangle\!_{C}^{2} \right)\nonumber\\
\!&&\!\!\!\!+\!  \lambda_{6}\!\left( \alpha \left| {{H}_{T}} \right\rangle\!_{A}^{1}\left| {{H}_{T}} \right\rangle\! _{B}^{2}\left| {{H}_{T}}
\right\rangle \!_{C}^{1}\!+\!\beta \left|{{V}_{T}} \right\rangle \!_{A}^{1}\left| {{V}_{T}} \right\rangle\!_{B}^{2}\left| {{V}_{T}} \right\rangle\!_{C}^{1} \right) \nonumber\\
\!&&\!\!\!\! +\!\lambda_{7}\!\left(\alpha \left| {{H}_{T}} \right\rangle\!_{A}^{1}\left| {{H}_{T}}\right\rangle \!_{B}^{1}\left| {{H}_{T}} \right\rangle\! _{C}^{2}\!+\!\beta\left|{{V}_{T}} \right\rangle \!_{A}^{1}\left| {{V}_{T}} \right\rangle\!_{B}^{1}\left| {{V}_{T}} \right\rangle\!_{C}^{2} \right)\nonumber\\
\!&&\!\!\!\!+\!\lambda_{8}\!\left( \alpha \left| {{H}_{T}} \right\rangle\!_{A}^{1}\left|{{H}_{T}} \right\rangle \!_{B}^{1}\left| {{H}_{T}} \right\rangle\!_{C}^{1}\!+\!\beta \left| {{V}_{T}} \right\rangle \!_{A}^{1}\left|{{V}_{T}} \right\rangle \!_{B}^{1}\left| {{V}_{T}} \right\rangle\!_{C}^{1} \right).\nonumber\\
\end{eqnarray}
\end{small}
Now, the three-photon system,  shared  by Alice, Bob, and Charlie, is in a predetermined entangled state as that sent by Susan, no matter in which output pathes the photons are detected. Therefore, the fidelity of the three-photon entanglement transmission is unity. Furthermore, the corresponding probabilities for different output paths depend on misalignment angles of the experimental reference frames of the parties, but the total success probability of the transmission is unity when ideal PCs are used.

Similarly, the scheme can also be extended to perform an alignment-free transmission for $N$-photon GHZ entanglements. First, a GHZ entangled $N$-photon system, sent by Susan, suffers a tag-operation $T_{V}$  on the vertical polarization component of each photon and completes the encoding process. Then,
when the $N$ photons arrive at the distant parties, they pass through the corresponding decoding circuit, which is constructed in the same structure as that for the two-photon entanglement transmission. The adverse effect, introduced by reference frames misalignment and  channel noises, can be totally compensated, since the $N$-photon system now evolves into a new state as
\begin{equation}
\begin{split}
{{\left| \varphi'  \right\rangle }_{AB...Z}}
\!=&\!\sum_{j,k,...l}{{{p}^A_{j}}{{p}^B_{k}}...{{p}^Z_{l}}}
\big(\alpha \left| {{H}_{T}}
\right\rangle _{A}^{{{t}_{j}}}\left| {{H}_{T}}
\right\rangle _{B}^{{{t}_{k}}}\cdot  \cdot \cdot \left| {{H}_{T}}
\right\rangle _{Z}^{{{t}_{l}}} \\&+\beta \left| {{V}_{T}}
\right\rangle_{A}^{{{t}_{j}}}\left| {{V}_{T}}
\right\rangle _{B}^{{{t}_{k}}}\cdot \cdot \cdot
\left| {{V}_{T}}  \right\rangle _{Z}^{{{t}_{l}}} \big),\label{eq-nghz}
\end{split}
\end{equation}
where the subscripts $i= A$, $B$, ..., and $Z$ represent photons that are described relative to the reference frames of Alice, Bob, ..., Zach, respectively.
$j,k,$ ..., $l~\in~\{0,1\}$.
${p^i_{0}}=cos{\theta' _{i}}$. ${{p}^i_{1}} =sin{{\theta' }_{i}}$.
The superscripts  $t_{0}=2$ and $t_{1}=1$ represent the photon outputs in path 2 and path 1, respectively. The $N$ receivers can share the desired $N$-photon GHZ polarization entanglement with each photon output in path 1 or path 2. Once the output paths are determined, the fidelity of the $N$-photon GHZ entanglement transmission is unity, and
the path-dependent success probabilities can be obtained from Eq.~(\ref{eq-nghz}). Moreover, the total success probability, in principle, is also unity when ideal PCs and linear optical elements are used.

\begin{figure}[!tpb]
\begin{center}
\includegraphics[width=7.0cm,angle=0]{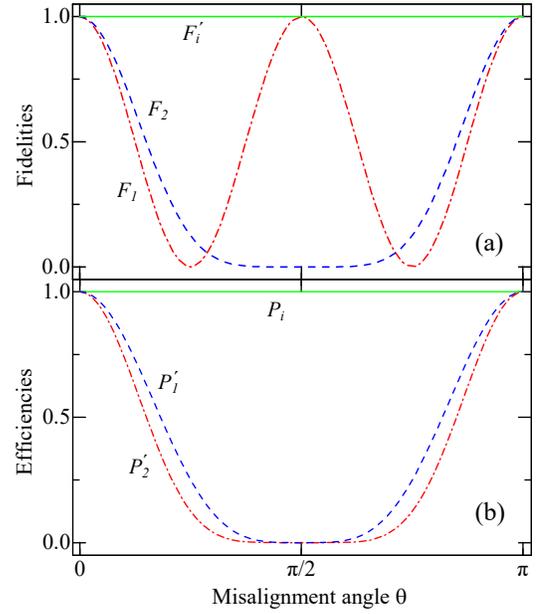}
\caption{(a) Fidelities versus misalignment angle $\theta$. $F_1$ and $F_2$ are  the fidelities of two- and three-party direct transmission, respectively. $F'_i=1$ is the transmission fidelity with our correction protocol; (b) Efficiencies versus misalignment angle $\theta$. $P'_1$ and $P'_2$ are the efficiencies of our two- and three-party transmission, respectively, and the direct transmission efficiency $P_i=1$. All misalignment angles are set to be the same ones with $\theta'_i=\theta$ for $i$=A, B, and C; $\alpha=1/\sqrt2$ and $\beta=-1/\sqrt2$.
}\label{fig6}
\end{center}
\end{figure}

\section{Performances of two- and three-party entanglement transmission}

As discussed in Secs.~\ref{Sec-2} and \ref{Sec-3}, the misalignment of experimental reference frames in distant communicating parties will introduce random rotations and lead to transmission errors, even when noise-free channels are used~\cite{Kalamidas,Kalamidas2006,Liapl2007,Shenglpl2014}. For a practical entanglement transmission, its fidelity and efficiency are two important parameters to measure its performance. In the alignment-free transmission protocol with linear optical elements, the fidelities of two- and three-party entanglement transmission are unity, by picking out the effective transmission with a delay $T$ on each photon. This transfers the infidelity introduced by the misalignment of experimental reference frames into an inefficiency.

The fidelities and efficiencies versus misalignment angles $\theta$ are shown in Fig.~\ref{fig6} (a) and (b), respectively. For simplicity, we assume all misalignment angles are the same ones with $\theta'_i=\theta$ for $i$=A, B, and C; $\alpha=1/\sqrt2$ and $\beta=-1/\sqrt2$. In addition, the channel photon loss, which increases exponentially with the transmission distance, is set to be \emph{zero} and the efficiency $P_i$ for two- and three-party direct transmission is unity. However, their respective fidelities $F_1$ and $F_2$ change significantly when the misalignment angle changes. In our protocol with linear optical elements, the fidelities $F'_i$ of two- and three-party entanglement transmission are always unity. However, our success efficiency $P'_1$ ($P'_2$) for two-party (three-party) entanglement transmission is dependent on the misalignment angle $\theta$, and it decreases when increasing $\theta$ for $\theta\leq\pi/2$, while it increases when increasing $\theta$ for $\theta\geq\pi/2$.

For the entanglement transmission with PCs,  we  have  assumed  that  each PC has  a perfect efficiency.  However, realistic PCs always  cause losses of photons passing through them with a finite transmission efficiency $\eta$.  The photon losses, induced by PCs, will finally decrease the efficiencies of the alignment-free two- and three-party entanglement transmission.
Fortunately, these losses have no effect on the fidelities of the transmission, since two PCs are used in each node and they are arranged in a symmetric structure, see Figs.~\ref{fig4} and~\ref{fig5}  for detail. For the two-party case, a photon pair passes through two PCs and the success efficiency is decreased from unity by a scale of $\eta^2$; For the three-party case, a photon triplet passes through three PCs and the corresponding efficiency is decreased to $\eta^3$. Currently, a PC with a transmission coefficient $\eta=98.8\%$ has been used for multi-photon quantum information processing \cite{PC-eta}, and $\eta$ can approach unity with advanced experimental technology. Therefore, our entanglement transmission with PCs, in principle, works in a deterministic way with a unity fidelity.\\

\section{Discussion and summary }

In long-distance quantum communication, it is of great importance to remove the requirement of reference-frame alignment, since it consumes infinite amount of classical resource to establish perfect aligned reference frames and the misalignment always leads to errors \cite{RFmis0}. We, therefore, propose two alignment-free polarization-entanglement transmission protocols:
one is based on passive linear-optical elements and the other one is based on active polarization modulation. In the first protocol, the distant communicating parties can share the desired entanglement in a probabilistic way, without the need to calibrate their reference frames with that of the sender. The infidelity, originating from either the channel noise or the reference-frame misalignment, is completely suppressed by a simple time-bin encoding and decoding process, and this is quite different from previous two- or multi-photon DFS-based protocols \cite{DFS2,DFS3,DFS4,Yamamoto,DFS6,Li2008}.

In the second protocol, the encoding process is the same as that used in the first protocol, while the  decoding process in each communicating party is  rearranged  with active polarization modulators. The parties can, in principle, share the desired entanglement with a unity fidelity and efficiency. This is much similar to protocols assisted by spatial-mode or OAM \cite{mdf1,mdf2,mdf3,Vallone2014}. However, our protocol is much simpler and more robust to channel noises, since our polarization-entanglement transmission is assisted by a stable time-bin DOF of  single photons other than the environment-sensitive spatial-mode or OAM.
Furthermore, PCs are well developed with near unity efficiency, and this makes our protocol more efficient when compared with that based on low-efficiency $q$-plate \cite{qp1,qp2}.

In summary, we have proposed two efficient alignment-free polarization-entanglement transmission protocols for entangled photons in faithful quantum communication and distributed quantum networks. These protocols can provide distant communicating parties with a desired polarization entanglement by new encoding-decoding processes. The infidelity originating from either channel noise and experimental reference frames misalignment is completely suppressed by either converting into a heralded loss or by active correction with high-efficiency polarization modulators. Furthermore, our protocols use only simple linear-optical elements and active Pockels cells, and it can significantly simplify quantum communication and distributed quantum networks.

\section*{ACKNOWLEDGMENTS}

This work was supported by the National Natural Science Foundation
of China under Grants No. 11674033, No.  11474026, and No. 11505007.

%

\end{document}